\documentstyle{elsart}

\input{psfig}

\begin{document}


\begin{frontmatter}

\title{Measurement of the D$_s$ Lifetime}

\collab{Fermilab E791 Collaboration}

\author[inst_9]{E.~M.~Aitala,}
\author[inst_1]{S.~Amato,}
\author[inst_1]{J.~C.~Anjos,}
\author[inst_5]{J.~A.~Appel,}
\author[inst_14]{D.~Ashery,}
\author[inst_5]{S.~Banerjee,}
\author[inst_1]{I.~Bediaga,}
\author[inst_8]{G.~Blaylock,}
\author[inst_15]{S.~B.~Bracker,}
\author[inst_13]{P.~R.~Burchat,}
\author[inst_6]{R.~A.~Burnstein,}
\author[inst_5]{T.~Carter,}
\author[inst_1]{H.~S.~Carvalho,}
\author[inst_12]{N.~K.~Copty,}
\author[inst_9]{L.~M.~Cremaldi,}
\author[inst_18]{C.~Darling,}
\author[inst_5]{K.~Denisenko,}
\author[inst_11]{A.~Fernandez,}
\author[inst_12]{G. F. Fox,}
\author[inst_2]{P.~Gagnon,}
\author[inst_1]{C.~Gobel,}
\author[inst_9]{K.~Gounder,}
\author[inst_5]{A.~M.~Halling,}
\author[inst_4]{G.~Herrera,}
\author[inst_14]{G.~Hurvits,}
\author[inst_5]{C.~James,}
\author[inst_6]{P.~A.~Kasper,}
\author[inst_5]{S.~Kwan,}
\author[inst_12]{D.~C.~Langs,}
\author[inst_2]{J.~Leslie,}
\author[inst_5]{B.~Lundberg,}
\author[inst_14]{S.~MayTal-Beck,}
\author[inst_3]{B.~Meadows,}
\author[inst_1]{J.~R.~T.~de~Mello~Neto,}
\author[inst_7]{D.~Mihalcea,}
\author[inst_16]{R.~H.~Milburn,}
\author[inst_1]{J.~M.~de~Miranda,}
\author[inst_16]{A.~Napier,}
\author[inst_7]{A.~Nguyen,}
\author[inst_3,inst_11]{A.~B.~d'Oliveira,}
\author[inst_2]{K.~O'Shaughnessy,}
\author[inst_6]{K.~C.~Peng,}
\author[inst_3]{L.~P.~Perera,}
\author[inst_12]{M.~V.~Purohit,}
\author[inst_9]{B.~Quinn,}
\author[inst_17]{S.~Radeztsky,}
\author[inst_9]{A.~Rafatian,}
\author[inst_7]{N.~W.~Reay,}
\author[inst_9]{J.~J.~Reidy,}
\author[inst_1]{A.~C.~dos Reis,}
\author[inst_6]{H.~A.~Rubin,}
\author[inst_9]{D.~A.~Sanders,}
\author[inst_3]{A.~K.~S.~Santha,}
\author[inst_1]{A.~F.~S.~Santoro,}
\author[inst_3]{A.~J.~Schwartz,}
\author[inst_4,inst_17]{M.~Sheaff,}
\author[inst_7]{R.~A.~Sidwell,}
\author[inst_18]{A.~J.~Slaughter,}
\author[inst_3]{M.~D.~Sokoloff,}
\author[inst_1]{J.~Solano,}
\author[inst_7]{N.~R.~Stanton,}
\author[inst_5]{R.~J.~Stefanski,}
\author[inst_17]{K.~Stenson}                                                   
\author[inst_9]{D.~J.~Summers,}
\author[inst_18]{S.~Takach,}
\author[inst_5]{K.~Thorne,}
\author[inst_7]{A.~K.~Tripathi,}
\author[inst_17]{S.~Watanabe,}
\author[inst_14]{R.~Weiss-Babai,}
\author[inst_10]{J.~Wiener,}
\author[inst_7]{N.~Witchey,}
\author[inst_18]{E.~Wolin,}
\author[inst_7]{S.~M.~Yang,}
\author[inst_9]{D.~Yi,}
\author[inst_7]{S.~Yoshida,}
\author[inst_13]{R.~Zaliznyak,}
\author[inst_7]{and C.~Zhang}

\address[inst_1]{Centro Brasileiro de Pesquisas F\'\i sicas, 
                 Rio de Janeiro RJ, Brazil}
\address[inst_2]{University of California, Santa Cruz, California 95064}
\address[inst_3]{University of Cincinnati, Cincinnati, Ohio 45221}
\address[inst_4]{CINVESTAV, 07000 Mexico City, DF Mexico}
\address[inst_5]{Fermilab, Batavia, Illinois 60510}
\address[inst_6]{Illinois Institute of Technology, Chicago, Illinois 60616}
\address[inst_7]{Kansas State University, Manhattan, Kansas 66506}
\address[inst_8]{University of Massachusetts, Amherst, Massachusetts 01003}
\address[inst_9]{University of Mississippi, University, Mississippi 38677}
\address[inst_10]{Princeton University, Princeton, New Jersey 08544}
\address[inst_11]{Universidad Autonoma de Puebla, Puebla, Mexico}
\address[inst_12]{University of South Carolina, Columbia, 
                  South Carolina 29208}
\address[inst_13]{Stanford University, Stanford, California 94305}
\address[inst_14]{Tel Aviv University, Tel Aviv, 69978 Israel}
\address[inst_15]{Box 1290, Enderby, British Columbia, V0E 1V0, Canada}
\address[inst_16]{Tufts University, Medford, Massachusetts 02155}
\address[inst_17]{University of Wisconsin, Madison, Wisconsin 53706}
\address[inst_18]{Yale University, New Haven, Connecticut 06511}

\begin{abstract}

We report the results of a precise measurement of the $D_s$ meson lifetime
based on $1662 \pm 56$ fully reconstructed $D_s \rightarrow \phi \pi$ decays,
from the charm hadroproduction experiment E791 at Fermilab.
Using an unbinned maximum likelihood fit, we measure the $D_s$
lifetime to be $0.518 \pm 0.014\pm 0.007$~ps.  
The ratio of the measured $D_s$ lifetime to the world average
$D^0$ lifetime~\cite{ref_pdg} is $1.25\pm0.04$.
This result differs from unity by six standard deviations,
indicating significantly different lifetimes for the $D_s$ and the $D^0$.


\end{abstract}

\end{frontmatter}


Precise measurements of the lifetimes of the weakly decaying
charm mesons are useful
for understanding the contributions of various weak decay mechanisms.
Despite the fact that they are all tied to the charm quark decay, the 
decays of the ground-state charm mesons can have different contributions 
from the four first-order processes 
(two spectator, W-annihilation, and W-exchange), 
and the lifetimes are, in fact, quite different~\cite{ref_pdg}:
\[
\tau(D^+):\tau(D^0):\tau(D_s) = 2.5:1:1.1
\]
The least well measured among these lifetimes is that of the $D_s$, which
is known only to about the 4\% level~\cite{ref_exp}.  
It is clear that there is a very large difference between the lifetime
of the $D^+$ and that of the $D^0$ or $D_s$. However, the current
measurements are only  suggestive of a difference 
between the lifetimes of the $D^0$ and the $D_s$:
\[
\frac{\tau(D_s)}{\tau(D^0)} = 1.13 \pm 0.04  \ \ \ \ \   
                    (\, 3 \sigma \ \mbox{difference from unity} \,).
\]
A more precise measurement of the lifetime of the $D_s$ could more clearly 
establish whether the lifetimes of the $D^0$ 
and $D_s$ are indeed different.

The large difference between the $D^+$ and 
$D^0$ lifetimes might be explained by a large contribution
from W-exchange in $D^0$ decays and/or large destructive interference
between $D^+$ decay amplitudes.
A difference between the $D^0$ and $D_s$ lifetimes could be due to 
phase space differences, color factors, W-exchange, W-annihilation, 
decay via $\tau$ lepton modes, or relativistic 
effects~\cite{ref_theory1,ref_theory2}.
The sensitivity of the $D^0$-$D_s$ lifetime difference to W-annihilation 
is noted in the paper by Bigi {\it et al.}~\cite{ref_theory2}.
However, spectator decays of the $D_s$ have one more strange quark 
than spectator decays of the $D^0$; hence, the Cabibbo-favored
hadronic final states have an extra kaon and correspondingly
smaller phase space. Table~\ref{tab_phase}
compares the phase space available for 2, 3, 4, and 5-body decays of
the two mesons. The table shows that phase space differences alone could
account for the currently observed lifetime differences.

In this letter, we report results of a high-statistics measurement 
of the $D_s$ lifetime based on data from the charm hadroproduction 
experiment E791 at Fermilab. 
(Unless otherwise specified, charge conjugate 
states are implicitly included in this letter.)
This measurement establishes more certainly a difference 
between the lifetimes of the $D^0$ and the $D_s$. 
We use only the decay mode $D_s \rightarrow \phi \pi$ since 
the $\phi$ mass constraint gives  
a $D_s$ sample with a very good signal-to-noise ratio. 
The reconstruction efficiency as a function of lifetime was determined from 
data for the Cabibbo-favored (CF) decay $D^+\rightarrow K^- \pi^+ \pi^+$ and 
from the ratio of the efficiency for $D_s \rightarrow \phi \pi$ to that for
$D^+ \rightarrow K^- \pi^+ \pi^+$  obtained from Monte Carlo (MC) simulations.
With this technique, most uncertainties
that arise from MC simulation should cancel.


E791~\cite{ref_detect} is a high statistics charm experiment
which acquired data at Fermilab during the 1991-1992 fixed-target run.
The experiment combined a
fast data acquisition system with an open trigger.
Over $2 \times 10^{10}$  events were collected with the 
Tagged Photon Spectrometer using a 500~GeV $\pi^-$ beam. 
There were five target foils
with 15~mm center-to-center separations: one 0.5~mm thick 
platinum foil followed by four 1.6~mm thick diamond foils. 
The spectrometer included
23 planes of silicon microstrip detectors (6 upstream and
17 downstream of the target), 2 dipole magnets, 10 planes of proportional wire 
chambers (8 upstream and 2 downstream of the target),
35 drift chamber planes, 2 multi-cell threshold \v{C}erenkov counters
that provided $\pi/K$ separation in the 6--60~GeV/$c$ momentum 
range~\cite{ref_ckv}, electromagnetic and hadronic calorimeters, and a
muon detector.


All criteria used to select candidate $D_s \rightarrow \phi \pi$ decays, 
with the exception of \v{C}erenkov identification requirements
and the $\phi$ mass cut,  
were chosen to maximize $S/\sqrt{S + B}$, where $S$ is the number of
$D^+ \rightarrow K^- \pi^+ \pi^+$  signal events in data, 
scaled to the level expected for $D_s \rightarrow \phi \pi$,
and $B$ is the number of background events from appropriate 
sideband regions of the $\phi \pi$ mass distribution.

To reconstruct the $D_s \rightarrow \phi \pi$ and 
$D^+ \rightarrow K^- \pi^+ \pi^+$ candidates, three-prong decay 
vertices with charge of $\pm 1$ were selected. 
All decay tracks were required to travel through at least one magnet
and be of good quality.
Decay vertices were required to be located outside the target foils,
and the significance of spatial separation from the primary vertex in the 
beam direction, $\Delta z/\sigma_{\Delta z}$,
where $\sigma_{\Delta z}$ is the error on the separation,  
was required to be at least 11.
The component of the $D$ momentum perpendicular to the line 
joining the primary and secondary vertices was required to be less 
than 0.35~GeV/$c$. The transverse impact parameter of the $D$ 
momentum with respect to the primary vertex was required to be less than
55~$\mu$m. Decay tracks were required to  pass closer to the secondary vertex 
than to the primary vertex, and the sum of the
squares of their momenta perpendicular to the $D$ 
direction was required to be larger than 0.3~(GeV/$c$)$^2$. 
The candidates in the decay mode 
$\phi \pi$ with $\phi \rightarrow K^+ K^-$ were also required to have 
$M(K^+ K^-)$ within $\pm 10$ MeV/$c^2$ of the $\phi$ mass.

For the decay $D^+ \rightarrow K^- \pi^+ \pi^+$,
the kaon was identified on the basis of charge alone
and no \v{C}erenkov identification cuts were needed.
For the two same-sign tracks in the decay $D_s \rightarrow \phi \pi$, 
the track with the highest \v{C}erenkov probability 
was assumed to be the kaon and the other track the pion. 
No further \v{C}erenkov identification cuts were applied, given the 
$\phi$ mass selection criteria.

The longitudinal and transverse position resolutions for the 
primary vertex were 
350~$\mu$m and 6~$\mu$m, respectively. 
The mean momentum of selected $D^+$ and $D_s$ mesons was 70~GeV/$c$. 
For the decay vertices of these mesons,
the transverse resolution was about 9~$\mu$m, nearly independent of 
the momentum, and the longitudinal resolution was about 360~$\mu$m at 
70~GeV/$c$ and worsened by 30~$\mu$m for every 10~GeV/$c$ increase
in momentum.

\begin{figure}
\vspace*{4 cm}
\centerline{\psfig{figure=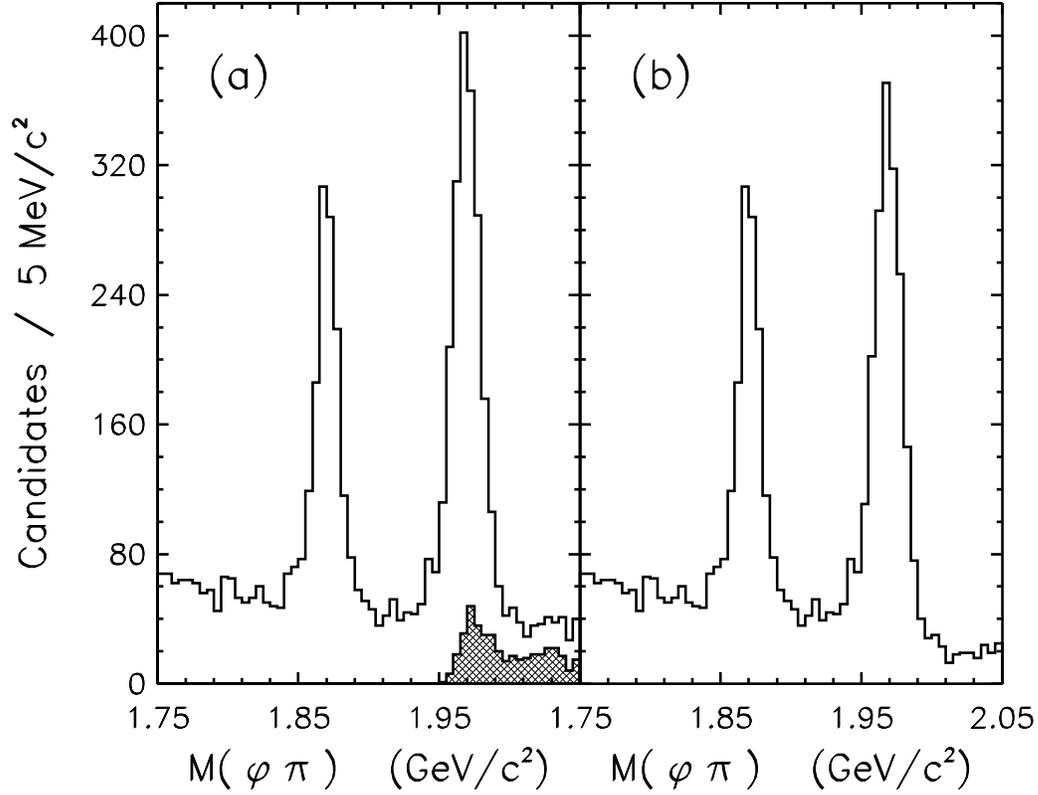,height=4.5in}}
\vspace*{2 cm}
\caption[]{The $\phi \pi$ mass distributions: (a) before excluding 
candidates with $M(K^- \pi^+ \pi^+)$ within 
$\pm 30$~MeV/$c^2$ of the $D^+$ mass,
which are shown as the hatched histogram; (b) after excluding  
candidates with $M(K^- \pi^+ \pi^+)$ within 
$\pm 30$~MeV/$c^2$ of the $D^+$ mass,
as described in the text.}
\label{fig_life_1}
\end{figure}

To eliminate any possible background due to reflections from 
$D^+ \rightarrow K^- \pi^+ \pi^+$ with a pion misidentified as a kaon,
which appear under and near the $D_s$ signal 
in the $\phi \pi$ mass plot, 
we excluded all $\phi \pi$ candidate events with $M(K^- \pi^+ \pi^+)$ 
within $\pm 30$ MeV/$c^2$ of the $D^+$ mass. 
The mass distributions for all $\phi \pi$ candidates
that survived our selection criteria 
are shown in Fig.~\ref{fig_life_1}.
The higher mass peak corresponds to $D_s \rightarrow \phi \pi$ 
while the lower peak corresponds to the singly 
Cabibbo-suppressed (SCS) decay $D^+ \rightarrow \phi \pi^+$.
The $\phi \pi$ mass distribution before 
$D^+ \rightarrow K^- \pi^+ \pi^+$ background subtraction 
is shown in Fig.~\ref{fig_life_1}(a), with candidates consistent with
$D^+ \rightarrow K^- \pi^+ \pi^+$ background shown 
as the hatched region. 
The $\phi \pi$ mass distribution after the reflection 
subtraction is shown in Fig.~\ref{fig_life_1}(b). 
The hatched region in Fig.~\ref{fig_life_1}(a) appears to be due to a small
$D_s$ signal and a linear background that turns on around 
1.95~GeV/$c^2$. The background in Fig.~\ref{fig_life_1}(b) is 
therefore modeled as a piecewise linear function with a discontinuity 
fixed at 1.95~GeV/$c^2$.  We have studied 
the sensitivity of our measured lifetime to the position of this
discontinuity and have found very little effect. We have included
a small systematic error in our total error to account
for the change in lifetime when the location of this discontinuity 
is varied from 1.95 to 1.99~GeV/$c^2$.


To extract the $D_s$ lifetime we performed a simultaneous unbinned 
maximum-likelihood fit to the distribution of mass and 
reduced proper decay time ($t^R$) 
of all $\phi \pi$ candidates that pass the selection criteria. 
The $t^R$ is defined as
\[
t^R = (L - L_{min})/c \beta \gamma,
\]
where $L$ is the distance between the secondary and primary 
vertices, and $L_{min}$ is the minimum $L$ value allowed by the cut 
on the separation between the secondary and primary vertices for this event.
We fit events in the ranges 
$1.75 < M(\phi \pi) < 2.05$~GeV/$c^2$ and $0.0 < t^R < 4.0$~ps.
From the fit we also measure the lifetime of the 
SCS decay $D^+ \rightarrow \phi \pi^+$.

The overall likelihood function is
\[
{\cal L} = \frac{e^{{-(N_{pred} - N_{obs})}^2/2N_{pred} }} 
{\sqrt{2 \pi N_{pred}}} \prod_i {\cal L}_i,
\]
with $N_{pred} = N_{D^+} + N_{D_s} + N_{BG}$, where $N_{D^+}$,
$N_{D_s}$, and  $N_{BG}$ are the fitted
number of $D^+ \rightarrow \phi \pi^+$,
$D_s \rightarrow \phi \pi$, and $\phi \pi$ background events, 
respectively.

The likelihood for each candidate is
\[
{\cal L}_i = S_{D^+}(m_i,t^R_i) + S_{D_s}(m_i,t^R_i) + B(m_i,t^R_i) 
\]
where
\begin{eqnarray} 
S_{D^+}(m_i,t^R_i) = F_{D^+} & \times & \frac{1}{\sqrt{2 \pi} 
 {\sigma_{D^+}}} e^{{-(m_{D^+}-m_i)}^2 / {2 \sigma_{D^+}}^2 } \nonumber\\
 & \times & R_{D^+} 
            \cdot f_{D^+}(t^R_i) \cdot e^{-t^R_i/\tau_{D^+}}, \nonumber
\end{eqnarray}
\begin{eqnarray} 
S_{D_s}(m_i,t^R_i) = F_{D_s} & \times & \frac{1}{\sqrt{2 \pi}
 {\sigma_{D_s}}} e^{ {-(m_{D_s}-m_i)}^2 / {2 \sigma_{D_s}}^2 } \nonumber\\
 & \times & R_{D_s} 
            \cdot f_{D_s}(t^R_i) \cdot e^{-t^R_i/\tau_{D_s}}, \nonumber
\end{eqnarray}
\begin{eqnarray} 
B(m_i,t^R_i) =  F_{BG} & \times & R_{BG} 
 \cdot (e^{-t^R_i/\tau_{BG_1}} + C e^{-t^R_i/\tau_{BG_2}})\nonumber \\
 & \times & \left\{ \begin{array}{ll}
 A_1 + S_1 m_i  & \ \ \ (\, \mbox{for} \ \ m_i < m_0 \,)  \\
 A_2 + S_2 m_i  & \ \ \ (\, \mbox{for} \ \ m_i \geq m_0 \,)
\end{array}. \right. \nonumber 
\end{eqnarray}

The functions $S_{D^+}(m_i,t^R_i)$, $S_{D_s}(m_i,t^R_i)$, and $B(m_i,t^R_i)$  
are the likelihood functions for $D^+ \rightarrow \phi \pi^+$, 
$D_s \rightarrow \phi \pi$, and $\phi \pi$ background, 
respectively.
The coefficients $F_{D^+}$, $F_{D_s}$, and $F_{BG}$ are the ratios
of the $D^+$, $D_s$, and the background, to the total number of events:  
$F_{D^+} = N_{D^+} / N_{pred}$, 
$F_{D_s} = N_{D_s} / N_{pred}$, and $F_{BG} = N_{BG} / N_{pred}$.
$m_i$ and $t^R_i$ are the mass and the reduced proper decay time 
of the $\phi \pi$  candidates.
$\sigma_{D^+}$ and $\sigma_{D_s}$ are the $D^+$ and $D_s$ 
mass resolutions. 
$\tau_{D^+}$ and $\tau_{D_s}$ are the $D^+$ and $D_s$ lifetimes.
The background in the mass distribution is described as two linear 
functions with slopes $S_1$ and $S_2$ which are determined 
from the fit, and $m_0$ is the position of the discontinuity in the 
background mass shape, which is fixed.
The terms $A_1$ and $A_2$ are obtained from the normalization and the 
size of the discontinuity, which is determined from the fit.
The coefficients $R_{D^+}$, $R_{D_s}$, 
and $R_{BG}$ are the normalizations of the 
fit functions over $t^R$ in the limits of the fit.

The functions $f_{D^+}(t^R_i)$ and $f_{D_s}(t^R_i)$ are the reconstruction
efficiency functions
of $t^R$, for $D^+ \rightarrow \phi \pi^+$ and $D_s \rightarrow \phi \pi$. 
These efficiency functions are determined from 
the CF $D^+ \rightarrow K^- \pi^+ \pi^+$ decay in
data, and from the ratios of MC efficiencies for 
the SCS $D^+ \rightarrow \phi \pi^+$ decay and 
the $D_s \rightarrow \phi \pi$ decay, separately, 
to that for $D^+ \rightarrow K^- \pi^+ \pi^+$ decay, as follows:
\begin{eqnarray} 
f_{D^+}(t^R_i) & = &
\epsilon^{Data}_{D^+ \rightarrow K \pi \pi}(t^R_i) \times
\frac{\epsilon^{MC}_{D^+ \rightarrow \phi \pi}(t^R_i)}
     {\epsilon^{MC}_{D^+ \rightarrow K \pi \pi}(t^R_i)}, \nonumber \\
& & \nonumber \\
f_{D_s}(t^R_i) & = &
\epsilon^{Data}_{D^+ \rightarrow K \pi \pi}(t^R_i) \times
\frac{\epsilon^{MC}_{D_s \rightarrow \phi \pi}(t^R_i)}
     {\epsilon^{MC}_{D^+ \rightarrow K \pi \pi}(t^R_i)}. \nonumber 
\end{eqnarray}
We determine the efficiency for $D^+ \rightarrow K^- \pi^+ \pi^+$ 
from data by comparing
the background subtracted $t^R$ distribution for 
$D^+ \rightarrow K^- \pi^+ \pi^+$ to an exponential function with 
the world average $D^+$ lifetime~\cite{ref_pdg}, 
$\tau(D^+) = 1.057 \pm 0.015$~ps. The uncertainty on the $D^+$ lifetime
contributes to our systematic error.
   
We prefer to determine the functions $f_{D}(t^R_i)$, 
$f_{D^+}(t^R_i)$ and $f_{D_s}(t^R_i)$, from data and the ratios of 
efficiencies from MC to reduce our dependency on MC. Most uncertainties
that arise from MC simulation should cancel in the ratios and, as a systematic
check, we also measure the $D_s$ lifetime using the functions 
$f_{D^+}(t^R_i)$ and $f_{D_s}(t^R_i)$ derived only from MC. 
We use the same cuts for the $\phi \pi$ and $K^- \pi^+ \pi^+$
data samples in order to minimize any bias in the efficiency ratios 
due to the  selection criteria.

The background in the $t^R$ distribution is parameterized as a sum of two 
exponential functions with lifetimes of $\tau_{BG_1}$ and $\tau_{BG_2}$, 
and a relative coefficient $C$. The parameters were determined 
from the $t^R$ 
distribution for the $D_s$ sidebands. As a systematic check
they were also allowed to float in 
the unbinned maximum-likelihood fit, with no significant effect on the 
measured $D_s$ lifetime; a small systematic error is included 
in our total error to account for this effect.

\begin{figure}
\vspace*{2 cm}
\centerline{\psfig{figure=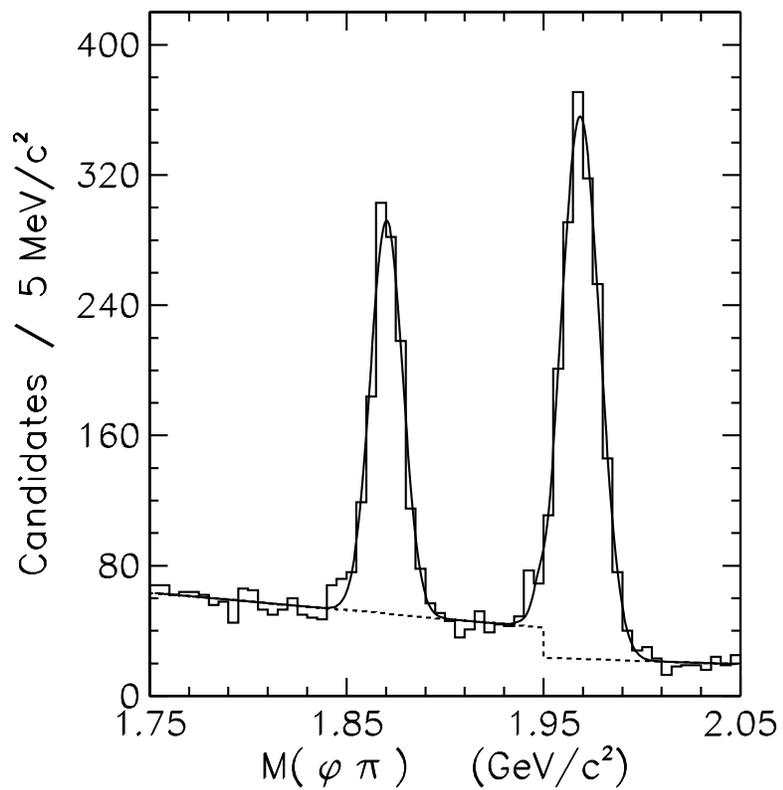,height=4.5in}}
\vspace*{2 cm}
\caption[]{The projection from the unbinned maximum-likelihood fit on the 
$M(\phi \pi)$ distribution, with the discontinuity in the linear
background fixed at 
1.95~GeV/$c^2$. 
The two peaks are the SCS $D^+$ signal with a fitted yield of $997 \pm 39$ 
events, and the $D_s$ signal with a fitted yield of $1662 \pm 56$ events,
as described in the text.}
\label{fig_life_2}
\end{figure}
\begin{figure}
\vspace*{2 cm}
\centerline{\psfig{figure=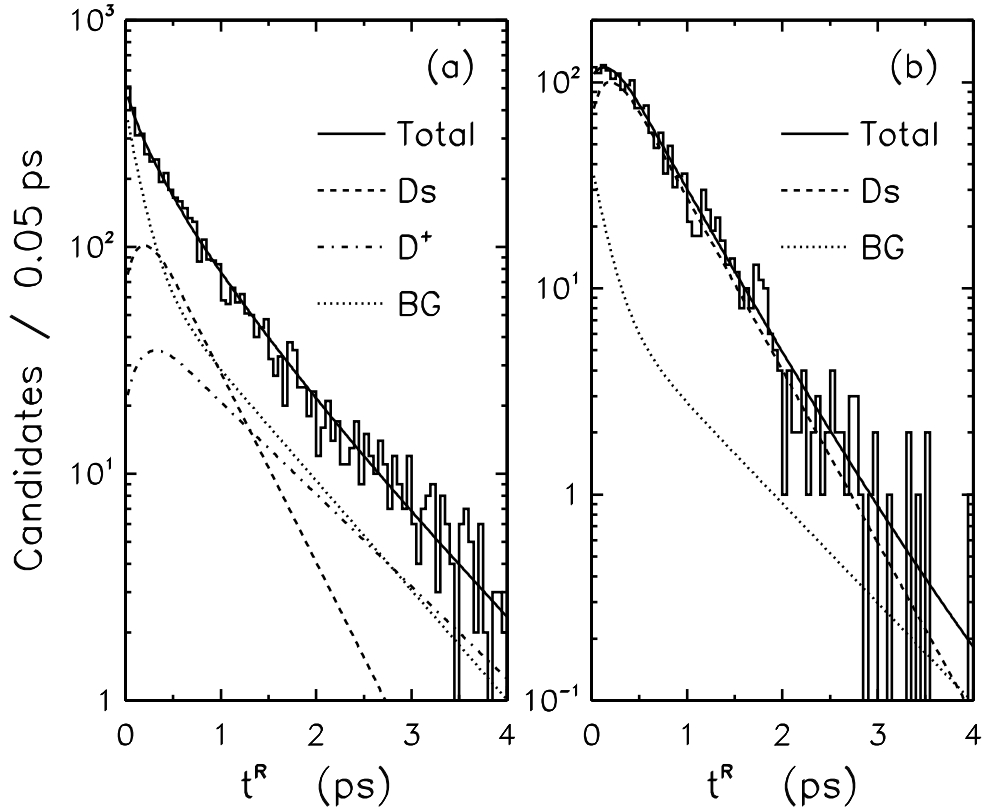,height=4.5in}}
\vspace*{2 cm}
\caption[]{The projection of the data and the best fit function
from the unbinned maximum-likelihood fit, 
on the reduced proper decay time distribution. (a) All events that pass 
the selection criteria  (histogram). 
Also shown are the total fit function (solid curve),
and the contribution to the fit from $D_s \rightarrow \phi \pi$ (dashed curve), 
$D^+ \rightarrow \phi \pi^+$ (dashed-dotted curve), and
$\phi \pi$ background (dotted curve).
(b) Events within $\pm 25$ MeV/$c^2$ of the $D_s$ mass (histogram). 
Also shown are the total fit function (solid curve),
and the contribution to the fit from $D_s \rightarrow \phi \pi$ (dashed curve)
and from $\phi \pi$ background (dotted curve),
as described in the text.} 
\label{fig_life_3}
\end{figure}

The projections of the data and the best fit function
from the unbinned maximum-likelihood fit on the $M(\phi \pi)$
and $t^R$ distributions for all events that pass the selection criteria
are shown in Figs.~\ref{fig_life_2} and  ~\ref{fig_life_3}(a), respectively. 
For the subset of events within $\pm 25$ MeV/$c^2$ of the $D_s$ mass,
the projections of these fit results on the $t^R$ distribution, 
scaled for the new mass range, are shown in Fig.~\ref{fig_life_3}(b).
This illustrates that our fit describes the data well for events in a narrow
window around the $D_s$ signal where the background is small.
The fit yields $1662 \pm 56$ 
$D_s \rightarrow \phi \pi$ signal events with a $D_s$ lifetime 
$\tau(D_s) = 0.518 \pm 0.014$~ps, and $997 \pm 39$
$D^+ \rightarrow \phi \pi^+$ signal events with a $D^+$ lifetime 
$\tau(D^+) = 1.065 \pm 0.048$~ps. The $D^+$ lifetime is in good agreement 
with current $D^+$ lifetime measurements, indicating that the MC
models the relative efficiency of $\phi \pi$ to $K^- \pi^+ \pi^+$ well.
The projections from the fit on the $t^R$ distribution for 
$D^+ \rightarrow \phi \pi^+$ and 
$D_s \rightarrow \phi \pi$ are not pure exponentials
because of the nonuniform detector efficiencies, which were parameterized by 
the functions $f_{D^+}(t^R_i)$ and $f_{D_s}(t^R_i)$ as described above.


Since ratios of MC efficiencies are used in the fit, most 
systematic errors cancel in the final $D_s$ lifetime measurement.  However,
some uncertainties remain.  The sources of these are summarized in 
Table~\ref{tab_syst}, and quantitative estimates of the errors are listed.  
In general, the estimates come from re-fitting after 
varying data selection criteria, parameters, and mass or $t^R$ regions used.
The total systematic error comes from adding the contributions 
in quadrature.

Other checks were performed such as fitting to
only a portion of the $t^R$ range, which would check our efficiency
functions across the $t^R$ range considered, 
and measuring the lifetime of the $D_s^+$ and $D_s^-$ separately.
Variations in results were found to be small compared to 
the statistical error.
Uncertainties due to target absorption and scattering of
charm secondaries are expected to be small due to the thinness of 
the target foils
and the requirement that decay vertices be located outside the target foils.
In addition, since we use the $D^+ \rightarrow K^- \pi^+ \pi^+$
data to correct for efficiency, any such 
effects should be accounted for. 

Finally, as a test of our fitting procedure, 
we fitted our $K^- \pi^+ \pi^+$
data to measure the $D^+$ lifetime using an unbinned maximum-likelihood 
fit similar to that used to fit the $\phi \pi$ data. 
The efficiency function 
for the $D^+ \rightarrow K^- \pi^+ \pi^+$ was determined
only from MC. The fit yielded $62651 \pm 337$  $D^+$ signal events,
shown in Ref.~\cite{ref_cpv},
with a measured $D^+$ lifetime that agrees with the world average
value~\cite{ref_pdg} to better than the 1.5\% error 
on the world average value. 


In conclusion, we have made a new precise measurement of the $D_s$
lifetime using $1662 \pm 56$ fully reconstructed 
$D_s \rightarrow \phi \pi$ decays.
The $D_s$ lifetime is measured using an unbinned maximum-likelihood fit
to be $0.518 \pm 0.014 \pm 0.007$~ps.
This value is about two standard deviations higher than the present world
average of $0.467 \pm 0.017$~ps~\cite{ref_pdg}. 
Using our result and the world average $D^0$ lifetime~\cite{ref_pdg},
we find the ratio of our $D_s$ lifetime to the $D^0$ lifetime to be
\[
\frac{\tau(D_s)}{\tau(D^0)} = 1.25 \pm 0.04  \ \ \ \ \ 
                    (\, 6 \sigma \ \mbox{difference from unity} \,)
\]
showing significantly different lifetimes for the $D_s$ and $D^0$.
This result may be used to
constrain the contributions of various decay mechanisms to charm 
decay and further refine our quantitative understanding of
the hierarchy of charm particle lifetimes.


We gratefully acknowledge the assistance of the staffs of Fermilab 
and of all the
participating institutions. This research was supported by the 
Brazilian Conselho Nacional de Desenvolvimento Cient\'\i fico e 
Technol\'{o}gio, CONACyT (Mexico), the Israeli Academy of Sciences
and Humanities, the U.S. Department of Energy, the U.S.-Israel
Binational Science Foundation and the U.S. National Science 
Foundation. Fermilab is operated by the Universities Research 
Association, Inc., under contract with the United States Department
of Energy.



\newpage

\vspace*{1in}

\begin{table}
\caption{Summary of the phase space (PS) factors, in arbitrary units,
and the phase space ratios, 
for 2, 3, 4, and 5-body decay modes of the $D^0$ and $D_s$.
The phase space factors are calculated for the final states shown,
with no intermediate resonances.} 
\label{tab_phase}
\bigskip
\begin{tabular}{l c | l c | c}
\hline
\multicolumn{2}{c|}{$D^0$} & \multicolumn{2}{c|}{$D_s$} &  \\
Decay Mode & PS & Decay Mode & PS & PS($D^0$)/PS($D_s$) \\
\hline
$K^- \pi^+$             & 1.45 & $K^+ \overline{K}^0$    & 1.36 & 1.07 \\
$K^- \pi^+ \pi^0$       & 1.70 & $K^+ K^- \pi^+$         & 1.32 & 1.29 \\
$K^- \pi^+ \pi^+ \pi^-$ & 0.537 & $K^0 K^- \pi^+ \pi^+$  & 0.247 & 2.17 \\
$\overline{K}^0 \pi^+ \pi^+ \pi^- \pi^-$ & 0.0548 &     
                            $K^+ K^- \pi^+ \pi^+ \pi^-$  & 0.0127 & 4.31 \\
\hline
\end{tabular} 
\end{table}

\begin{table}
\caption{Summary of the systematic uncertainties assigned to 
the $D_s$ lifetime, as described in the text.} 
\label{tab_syst}
\bigskip
\begin{tabular}{l c}
\hline
Source of Uncertainty & Systematic Error \\
\hline
$D^+$ lifetime uncertainty~\cite{ref_pdg}                  & 0.004~ps \\
Size of $D^+ \rightarrow K^- \pi^+ \pi^+$ reflection       & 0.003~ps \\
Data selection criteria                                    & 0.003~ps \\
Likelihood efficiency functions, $f_{D}(t^R_i)$            & 0.003~ps \\
$t^R$ parameterization of background                       & 0.002~ps \\
Position of step in background function                    & 0.002~ps \\
Monte Carlo production model                               & 0.002~ps \\
\hline
Total systematic error                                     & 0.007~ps \\
\hline
\end{tabular} 
\end{table}


\end{document}